\documentclass[letterpaper]{article} 
\pdfoutput=1
\usepackage{pdfpages}
\usepackage{aaai23}  
\usepackage{times}  
\usepackage{helvet}  
\usepackage{courier}  
\usepackage[hyphens]{url}  
\usepackage{graphicx} 
\usepackage[numbers]{natbib}  
\usepackage{caption} 
\usepackage{algorithm}
\usepackage{algorithmic}

\usepackage{newfloat}
\usepackage{listings}
\DeclareCaptionStyle{ruled}{labelfont=normalfont,labelsep=colon,strut=off} 
\floatstyle{ruled}
\newfloat{listing}{tb}{lst}{}
\floatname{listing}{Listing}
\title{Merging AI Incidents Research with Political Misinformation Research: Introducing the Political Deepfakes Incidents Database}
\author{
    Christina P. Walker,\textsuperscript{\rm 1}\\
    Daniel S. Schiff,\\
    Kaylyn Jackson Schiff
}
\affiliations{
    \textsuperscript{\rm 1}Purdue University, Department of Political Science\\

    walke667@purdue.edu
}

\begin{document}

\maketitle

\begin{abstract}
This article presents the Political Deepfakes Incidents Database (PDID), a collection of politically-salient deepfakes, encompassing synthetically-created videos, images, and less-sophisticated `cheapfakes.' The project is driven by the rise of generative AI in politics, ongoing policy efforts to address harms, and the need to connect AI incidents and political communication research. The database contains political deepfake content, metadata, and researcher-coded descriptors drawn from political science, public policy, communication, and misinformation studies. It aims to help reveal the prevalence, trends, and impact of political deepfakes, such as those featuring major political figures or events. The PDID can benefit policymakers, researchers, journalists, fact-checkers, and the public by providing insights into deepfake usage, aiding in regulation, enabling in-depth analyses, supporting fact-checking and trust-building efforts, and raising awareness of political deepfakes. It is suitable for research and application on media effects, political discourse, AI ethics, technology governance, media literacy, and countermeasures.

\end{abstract}

\section{Background and Motivation}
\subsection{Recent Advances and Political Implications}
Deepfake generation capabilities have evolved significantly in recent years, with the technology now capable of creating hyper-realistic content with minimal expertise \citep{barariPoliticalDeepfakesAre2021}. This includes not only synthetic text, but also audio, images, and videos, making it increasingly difficult for individuals to reliably distinguish between real and fake content \citep{grohDeepfakeDetectionHuman2022, ranaDeepfakeDetectionSystematic2022}. While deepfakes and other synthetic media, including simpler `cheapfakes,' serve diverse purposes from entertainment and education to fraud, a class of particularly concerning uses are politically-oriented deepfakes. Recent instances include the use of deepfakes to critique political figures such as Donald Trump, Joe Biden, Emmanuel Macron, Nancy Pelosi, Vladimir Putin, and Volodymyr Zelenskyy.\footnote{These deepfakes are included in the PDID, which can be found at: http://bit.ly/pdid.} Deepfakes have been leveraged to influence voter turnout \citep{christopherWeVeJust2020}, accuse politicians of sexual scandals \citep{cookDeepFakesCould2019}, and even influence geopolitical events like the Russian invasion of Ukraine \citep{simoniteZelenskyDeepfakeWas2022}.

Additionally, political campaigns need to be aware of the potential use of deepfakes. 
Candidates in the 2024 U.S. presidential primary races have already created campaign ads involving deepfakes \citep{nehamas_desantis_2023}. This is especially problematic as broadcast networks cannot refuse to run a campaign ad, as The Communications Act of 1934 says they have ``no power of censorship" over any ``legally qualified candidate" for office \citep{montanaro_truth_2022}. In response, the Federal Election Commission voted unanimously to pursue evaluating the rules around AI use in campaigns \citep{klarDemocratsLaudFederal2023}. 

Consequences can be far-reaching. For example, a fabricated video depicting a Pentagon bombing caused a dip in the U.S. stock market \citep{sorkin_i-generated_2023}. In another instance, mere accusation of a deepfake helped facilitate a coup in Gabon \citep{cahlan_analysis_2020}. Despite efforts by leading artificial intelligence (AI) providers to curb the creation of politically harmful content, researchers have demonstrated the ease with which mainstream tools can be bypassed. For example, Midjourney prompts like ``evil politicians grinning, sad children, Comet pizza shop'' and ``George Floyd realistic robbing a Wal-Mart'' \citep{albaMidjourneyEasilyTricked2023}, referencing a political conspiracy theory and a contentious high-profile police incident in the U.S., respectively, can be used to generate fake images potentially capable of stoking real political divisions. Consequently, scholars have urged that deepfakes could extend the harms of mis- and disinformation. This includes influencing elections, furthering democratic backsliding, undermining trust in media and government institutions, fostering pervasive uncertainty, inciting intolerance and violence, undermining public health, providing new tools for disinformation campaigns, harming psychological security, and empowering authoritarian governments \citep{sharmaExaminingMotivationsSharing2023, goldsteinForecastingPotentialMisuses2023, vaccariDeepfakesDisinformationExploring2020, schiffLiarDividendImpact2023}.

\subsection{Policy Responses}
International, national, and subnational entities, along with private sector actors, researchers, civil society groups, and standards organizations have already started to develop policies and regulations to address these concerns. Responses range from deepfake detection methods and watermarking to policy taskforces, platform regulation, media literacy efforts, and outright bans \citep{ranaDeepfakeDetectionSystematic2022, partnershiponaiPAIResponsiblePractices2023}. The European Union, for instance, has been advancing efforts to regulate the use of AI systems including deepfakes through risk assessments and transparency requirements \citep{europeancommissionProposalRegulationEuropean2021}. China has enacted regulations of deepfakes with criminal penalties \citep{haoChinaPioneerRegulating2023}. The U.S. has also considered a deepfake task force and secured a commitment from leading AI companies to watermark AI content, furthered by a 2023 Executive Order \citep{ stokolsWhiteHousePress2023}, and Congress has proposed the DEEPFAKES Accountability Act. Similarly, states and cities are also exploring legislative measures to curb the misuse of deepfakes, with nine U.S. states enacting regulation as of 2023 in lieu of federal action \citep{williamsExploringLegalApproaches2023}. 

Although some deepfake databases have been developed, such as the FakeCatcher database, AI-Powered Video Archive, and the Deepfake Detection Challenge Dataset, these are largely aimed at advancing technical detection methods and not comprehensive across the spectrum of politically and publicly salient deep- and cheapfakes, making systematic analysis challenging. The growing legislative concern along with the increasing sophistication of deepfakes underscores the need for a comprehensive incidents database, which aims to provide the means for valuable insights into the prevalence and impact of political deepfakes.   

We thus introduce the Political Deepfakes Incidents Database (PDID). The fundamental goal of the database is to provide a publicly available resource to advance research, practice, and governance efforts surrounding deepfakes. Journalists, fact-checking organizations, media literacy educators, and members of the public can utilize the database to evaluate the veracity of specific instances or identify broader trends of concern. In addition, the PDID can assist in understanding the effectiveness of watermarking and detection techniques, media literacy efforts, and other policy initiatives on deepfake dissemination and impact. 

\section{Creating a Political Deepfakes Database}
\subsection{Disciplinary Gaps in Deepfake Research} 

With the increased usage of generative AI in politics, there is a corresponding need to advance research into political deepfakes. Much research to date has focused on whether deepfakes are convincing to viewers \citep{barariPoliticalDeepfakesAre2021}, whether they are persuasive in terms of changing beliefs or attitudes \citep{schiffLiarDividendImpact2023, wittenbergMinimalPersuasiveAdvantage2021}, and whether they can be detected \citep{grohDeepfakeDetectionHuman2022}. Yet many of these studies take place in idealized laboratory studies using research-created content, undermining our capacity to understand real-world behavior and effects. Thus, observational data are needed to assess the prevalence and impact of `real' deepfakes with heightened political salience that \textit{have gained public attention}.

Related but broader concerns about the potential harms of AI have led to the initiation of a research agenda around AI incidents. Examples include the AI Incidents Database which now features more than 500 incidents as of 2023 \citep{mcgregorPreventingRepeatedReal2021}, the Awful AI Database \citep{daoDaviddaoAwfulaiAwful2022}, and the AI Vulnerability Database \citep{avidAIVulnerabilityDatabase2023}. Reporting certain kinds of incidents may soon become a requirement by international regulators, though potentially limited to incidents caused by products and services of companies, rather than adversarial actors or private individuals. Indeed, one challenge identified by this research is the existence of many more reported `issues' where harm is not as easily detectable \citep{mcgregorIndexingAIRisks2022}. 

Political misinformation is one such case, where serious consequences are possible, such as social upheaval, political instability, collapses in institutional trust, and ascendance of unaccountable or authoritarian leaders. However, the pathway to these harms is difficult to identify causally; individuals may be exposed to large amounts of content over time including messages and counter messages, making it difficult to identify single deepfakes as unitary sources of harm. Information-seeking behavior and attitudinal change may differ across demographic subgroups, such as across regions or groups with different levels of digital literacy. Harms may thus result from pervasive inputs acting through complex causal chains, with feedback effects and heterogenous effects rendering analysis difficult. Nevertheless, given the growing importance of this research as misinformation increases and democracy backslides, determining strategies for identifying and mitigating deepfake-related harms is critical. The growing toolkit of misinformation researchers can be useful here to support work being done on AI incidents collection and analysis, currently primarily by technical researchers or AI generalists. For example, scholars in political science, communications, and related fields have developed and tested theories and models to evaluate issues like:

\begin{enumerate}
    \item How actors or institutions are incentivized to promote misinformation/disinformation
    \item How misinformation is shared, promoted, and accessed through various channels
    \item How different types and characteristics of policy frames or narrative strategies can strategically shape views of issues, politicians, or social groups
    \item How individual and group political psychology, including in-group and out-group behavior, influence the efficacy of misinformation
    \item How the timing of messaging and counter-messaging leads to persistence effects or decay effects as contrasting messages compete or are reinforced
    \item How observational, quasi-experimental, and experimental evidence can be mustered to test the validity of theories, including studies of the efficacy of policy changes, mitigation efforts, and choices around AI system design
\end{enumerate}

To enable effective research on political deepfakes and mitigations then, information about the prevalence and nature of deepfakes must be curated and aligned with theoretical elements that a broader set of researchers can draw on to advance practical knowledge and associated action.

\subsection{Purpose of the Political Deepfakes Incidents Database (PDID)}

To foster these goals, this article introduces the Political Deepfakes Incidents Database (PDID). The goal of this database is to create a comprehensive and interoperable collection of political deepfakes and cheapfakes, beginning with videos and images\footnote{We omit audio-only content initially in order to set a realistic scope for the first phase of the PDID. Individuals tend to believe visual content due to a strong ``realism heuristic'' \citep{vaccariDeepfakesDisinformationExploring2020}, so we focus on the misinformation that arguably has the largest potential impact on trust in the informational environment. We may incorporate modalities like audio in the future.} primarily from the U.S.\footnote{We also initially limit our scope to the U.S., with intentions to scale the database gradually, likely relying on the on-the-ground expertise of partners in other countries and regions.}, to facilitate research on their usage in politics. The PDID is motivated by and aims to draw on the informational needs of misinformation researchers and practitioners---in media studies, political communication, artificial intelligence, public policy, and political science, amongst others---to enable the study of the prevalence, nature, and impact of political misinformation. By incorporating deepfakes, their content and context, real-world connections, and related aspects of politics and policy coded by researchers, the database aims to provide a valuable resource for understanding the prevalence, trends, and potential impact of political deepfakes. Critically, the PDID enables empirical studies helping to fill one of the key gaps in current research: the lack of causal evidence linking the use of political deepfakes to real-world outcomes. The database thus holds potential for various stakeholders, including policymakers, researchers, journalists, fact-checkers, and the public. 

Policy leaders can benefit from this database by gaining insights into the scope and scale of political deepfake manipulation. Understanding how deepfakes are used in both international and domestic politics can help policymakers formulate targeted regulations and measures to address the challenges posed by this technology while safeguarding democratic processes and national security. Scholars and analysts can utilize this database to conduct in-depth analyses and investigations into the patterns and dissemination of political deepfakes. By studying the content, context, and impact of these manipulated materials, researchers can enhance our theoretical understanding of the societal, political, and psychological implications of deepfakes. 

For example, in the field of communication, the PDID can provide insights into how deepfakes are crafted and disseminated and how they influence public discourse. Information technology practitioners can use the database as a resource for developing and testing new algorithms for detecting deepfakes. Political scientists can investigate the role of deepfakes in shaping political narratives, influencing elections, and destabilizing democratic processes. Fact-checking organizations and journalists can use the PDID to cross-reference the authenticity of content and to report on trends. Finally, the PDID can help educate the public about the existence and potential dangers of political deepfakes.

\section{Methodology}
\subsection{Data Collection and Codebook Development}
Data collection began in June 2023 and is ongoing, while the PDID currently includes deepfakes incidents dating back to 2017. The database is hosted on the platform AirTable,\footnote{The database is available at: http://bit.ly/pdid.} which the research team used to manually code the variables described below for each deepfake incident. Data are primarily collected from English language social media posts and popular news websites given our focus on deepfakes that have reached the public eye.  
To capture associated data, we begin with these news sources reporting on prominent deepfake incidents and use a snowball sampling technique, looking at linked articles and social media posts. 

To create the structure for the PDID, we began by collecting an initial sample of prominent political deepfakes, which we define as deepfakes featuring or commenting on prominent political figures, both individuals and groups, as well as politically important social issues and events.  
We then iteratively developed a codebook to guide data collection and coding for the PDID, for coders, and later analysis and interpretation of the database. We determined the relevant categories based on the data we collected for the initial sample, and we identified various theoretically important indicators based on recent political science and communication research and external feedback. To improve interoperability and enhance rigor towards greater impact, we consulted existing and forthcoming coding taxonomies developed by the AI Incidents Database, Center for Security and Emerging Technology, and AI Vulnerabilities Database.\footnote{We thank these teams for sharing their insights on the PDID.} 

The criteria that we used to develop the codebook are as follows. First, the variables needed to parse features of deepfakes with sufficient variation and granularity to enable meaningful study (e.g., the harm type depicted). Second, the coding choices for the incidents needed to be reliable and replicable (e.g., an original source could be identified with sufficient confidence based on information in the deepfake, key social media post, or news article). Finally, certain variables were identified because they held potential theoretical importance for misinformation researchers (e.g., the framing of a deepfake). We thus divided the codebook into categories including metadata, verification data, entity descriptors, and social media and sharing information, along with variables surrounding context and content, real-world connections, and theoretical codes related to politics and policy.

\subsection{Database Structure and Variables}

As noted, we incorporated variables from the AI Incidents Database and CSET taxonomy that we thought were both relevant and analytically tractable, removed inapplicable items, and added breadth and depth where appropriate. For example, while CSET's taxonomy has a special focus on identifiable and often physical harms for substantive and methodological reasons, the PDID codebook was designed to capture theoretically relevant aspects of political deepfakes and thus was modified to capture less `tangible' political and social harms. 

Table \ref{tab:variables_overview} provides a simplified overview of the variables included in the PDID, and 
the online Appendix\footnote{Available online at DOI 10.17605/OSF.IO/FVQG3.} provides provides full variable lists and definitions, which we expect to evolve over time. Below, we provide detail on the motivation and construction of some key variables of interest. 

\textbf{Metadata:} For metadata, we capture information such as the fake itself (e.g., deepfake or cheapfake, video or image), whether the incident has been verified as fake or authentic, the level of social media engagement (e.g., number of likes), and when available, information on the type, name, and location of the target and sharer entities. 

\begin{table}[!t]
    \centering
    \resizebox{\columnwidth}{!}{%
    \fontsize{8pt}{10pt}\selectfont 
    \begin{tabular}{|p{2.3cm}|p{4.2cm}|}
        \hline
         \textbf{Descriptors} &  URL, File, Media format, Summary \\
         \hline
         \textbf{Social Media} & Views, Likes, Comments, Shares \\
         \hline
         \textbf{Source} & Name, Entity type, Location \\
         \hline
         \textbf{Sharer} & Name, Occupation, No. of followers\\
         \hline
         \textbf{Target(s)} & Name, Entity type, Response (if any)\\
         \hline
         \textbf{Verification} & Presentation as real or fake, Visual markings of false content, Watermark/label, External sources \\
         \hline
         \textbf{Context \& Content} & Text from post, Context from post and comments, Harm type depicted (e.g., violence or discrimination), Communication goal (e.g., satire or reputational harm)\\
         \hline
         \textbf{Real-world \newline Connections} & Context (e.g., scandal, election, or war), Reported real-world harms (e.g., evidence of actual violence or political interference) \\
         \hline
         \textbf{Politics \& Policy} & Policy sector, Framing, Narrative \\
         \hline
    \end{tabular}
    }
    \caption{Overview of Variables in PDID}
    \label{tab:variables_overview}
\end{table}

\textbf{Verification:} We collect information regarding whether the images and videos in the PDID are authentic or fake. We code 1) whether the sharer presented the image or video as real or fake, and 2) any external verification of veracity. Coders also identify visual markings that could indicate content is synthetic (e.g., extra fingers, blurred faces).

\textbf{Context and Content:} We collect information regarding the context and content of the deepfakes, using text or transcripts from the image or video itself as well as from the primary associated post and surrounding comments. We further identify \textit{communication goals} for each deepfake, encompassing a range of objectives from political interference, emotional and psychological harm, and reputational damage to satire, entertainment, and education. That is, while deepfakes are often associated with spreading false information, our coding strategy acknowledges their potential for goals like education or raising awareness \citep{schegloffWhoseTextWhose1997, poriaContextDependentSentimentAnalysis2017}. Additionally, we consider potential harms that political deepfakes might \textit{depict}. Unlike existing frameworks focusing on physical harm, we develop a harm type typology that encompasses a broader range of harms. In particular, we identify not only whether political deepfakes depict harms associated with violence or property damage, but also whether they depict human rights violations, discrimination, or even environmental harms. Communication goals and harm types are not mutually exclusive, and we aim for a comprehensive understanding of deepfakes' multifaceted dangers. Determining communication goals and harm types involves subjectivity and inference, but is crucial for understanding these complex social phenomena. 

\textbf{Real-world Connections:} Next, the PDID aims to connect political deepfakes to \textit{real-world events and harms} to better understand the motivation for their creation and spread. We identify whether each deepfake is associated with contemporaneous events such as scandals, crimes, elections, or wars. We also code whether there is potential evidence, based on news reports, of actual harms tied to the deepfakes. For example, as described previously, a fake image of a bombing at the Pentagon caused financial harm through a dip in the stock market. We thus identify whether political deepfakes have caused harms related to financial loss, political interference, discrimination, human rights, violence, property damange, or the environment. For transparency, coders provide information regarding the evidence for their inferences, such as links to external reporting. 

\textbf{Politics and Policy:} Given that the PDID includes political deepfakes, we also devote particular attention to factors related to politics and policy. We categorize the \textit{policy sectors} most relevant to each deepfake using definitions from the CSET policy sector typology and the U.S. Census Bureau \citep{bureau_economic_2022}. We distinguish between sectors like medicine, transportation, and public safety/military, and we allow deepfakes to fall into multiple policy sectors. This helps to identify intended targets and potential impacts, revealing patterns in sectors most affected by or vulnerable to deepfakes. Next, we extend \textit{framing} research, which typically focuses on the media \citep{matthesWhatFrameContent2009} and framing efforts by actors like advocacy groups \citep{debruyckerFramingAdvocacyResearch2017}, to deepfakes and the actors sharing them. The framing of deepfakes refers to the perspectives, themes, or underlying ideas that shape how the manipulated content is communicated. We draw on \citet{semetko2000FramingEuropeanpolitics}'s typology of human interest, conflict, economic consequences, morality, and responsibility frames, which helps to uncover underlying motivations, such as intentions to exacerbate conflicts or assign blame. Finally, we identify aspects of the \textit{policy narrative} represented within each deepfake. This includes who or what the deepfake portrays as the hero, villain, plot, and moral, building on research on the Narrative Policy Framework (NPF) in the policy process literature \citep{shanahanNarrativePolicyFramework2011}. The study of narratives allows for investigations into how deepfakes impact individuals' understanding of policy actors, underlying responsibility, and social problems and solutions.

\section{Conclusion}

\textbf{Opportunities for Research and Practice:}
The PDID provides numerous opportunities to address harms and advance governance efforts related to political deepfakes. From a research perspective, the database is well-suited for longitudinal studies, surveys, experiments, and descriptive analyses. Analysts focused on political actors, institutions, and the political economy of misinformation can examine which actors are employing deepfakes and for what purposes. Further, the PDID can be used to analyze the prevalence of different frames or narratives used in political deepfakes, shedding light on the strategies employed to shape public perceptions. The database can thus facilitate research on how exposure to political deepfakes influences public opinion, trust in media, and other key aspects of political behavior, such as voting. 

The database can also advance research on ethical considerations surrounding AI, the creation and distribution of political deepfakes, and legal implications related to privacy, defamation, image-based sexual violence, and propaganda. Researchers and practitioners can use the database to investigate how political deepfakes impact public discourse, democratic processes, and civic engagement. Importantly, within each of the domains described above, researchers could also conduct comparative studies across different political contexts and demographic subgroups to explore how the use of and responses to political deepfakes vary. 

The PDID can also support researchers, practitioners, and policymakers working on AI policy and technology governance. For example, it can enable exploration into how policy changes or advancements in deepfake technologies influence the distribution and content of political deepfakes over time. Did the public release of large language models like ChatGPT change how AI-generated political content is disseminated? Can the addition of watermarks to deepfakes successfully increase detection or decrease their spread? Can enhanced transparency requirements, incident reporting, fact-checking, or auditing measures counteract harms from deepfake proliferation or improve public resilience?
 
Importantly then, the PDID can aid in developing and evaluating countermeasures, fact-checking efforts, detection techniques, and policy actions by AI developers, social media companies, standards organizations, and regulators. This includes informing the development of media or digital literacy initiatives to equip individuals with the skills to identify and critically assess political deepfakes. A central purpose of the PDID, in linking AI incidents research with misinformation research, is to help actualize an applied research agenda to address these pressing societal challenges.

\textbf{Future Directions:} Nonetheless, realizing these goals will require significant effort to enhance, govern, and sustain the PDID. Key goals and associated challenges include:
\begin{enumerate}
    \item \textbf{Maintaining and sustaining the database}. The PDID is intended to adopt FAIR principles (Findable, Accessible, Interoperable, and Reusable), including preserving an associated downloaded file for each deepfake. Pending consideration of increased ethical risks, the PDID will be an open-access resource, easily available for researchers, policymakers, journalists, and citizens. This requires sustainable infrastructure, funding, and staffing.
    \item \textbf{Enabling public incident reporting}. This will facilitate scaling, shared governance, and attention to incidents that could otherwise be neglected, and requires infrastructures for both reporting and vetting reports.
    \item \textbf{Promoting ecosystem interoperability}. We aim to make the PDID interoperable with other AI incident and vulnerability reporting tools. Ongoing conversations with AIID and AVID, and future engagement with government-sanctioned incident reporting, will improve the impact and reach of the PDID and related efforts.
    \item \textbf{Expanding regional and linguistic coverage}. Language and regional coverage limitations represent a significant gap in the current PDID, exacerbating a dilemma that individuals in low-income regions can suffer from both more misinformation and less mitigation. The PDID will benefit from efforts to extend data collection to other platforms, communication methods, countries, and languages. This will require careful scaling, local expertise, political sensitivity, and multilayered governance.
    \item \textbf{Developing shared governanc}e. The PDID will only be effective if it is actually used by researchers and other key practitioners. We will thus continue efforts to engage with journalists, fact-checking organizations, policymakers, researchers, and civil society groups. Governance mechanisms could include, for example, a steering committee and ongoing forums for feedback to enhance the utility of the PDID for different stakeholders.
    \item \textbf{Expanding dataset coverage}. As policy developments and regulations impact how deepfakes are produced and disseminated (e.g., through the inclusion of watermarks or formal verification efforts), we will iterate on the breadth and depth of the coding categories in the PDID to provide a more robust, fine-grained, and comprehensive database. We will also explore automated techniques to improve coverage and efficiency compared to manual coding. Finally, expanding the PDID to encompass other media formats like audio may be socially important.
    
\end{enumerate}

\subsubsection{Ethical Statement.} 
Curating and storing deepfakes has the potential to increase access to harmful content. However, we believe that the PDID's record of historical and verified (when possible) deepfakes is unlikely to be useful for adversarial purposes above and beyond existisg tools, while offering new benefits for research, verification, and mitigation. Given concerns about psychological harms from exposure to deepfakes, the PDID excludes highly incendiary sexual or violent content. We will continue to weigh the benefits and risks of open access data on political deepfakes in decisions about the public release and presentation of the PDID.

\section{Acknowledgments}
The authors would like to thank the Center for Security and Emerging Technology (CSET) for access to their harms taxonomy and codebook on AI Incidents, which supported the development of the PDID codebook. We are grateful to the AI Incidents Database (AIID) and AI Vulnerabilities Database (AVID) teams for their insights and feedback on interoperability. We also would like to thank J.P. Messina for helping to develop the codebook and broader project agenda, and are grateful to our undergraduate research assistants Paige Carter, Sharvani Kondapally, and Shreya Venkat.

\label{sec:references}
\bibliography{references, References-Daniel}

\begin{thebibliography}{33}
\providecommand{\natexlab}[1]{#1}

\bibitem[{Alba(2023)}]{albaMidjourneyEasilyTricked2023}
Alba, D. 2023.
\newblock Midjourney {{Is Easily Tricked Into Making AI Misinformation}}, {{Study Finds}}.
\newblock \emph{Bloomberg.com}.

\bibitem[{{AVID}(2023)}]{avidAIVulnerabilityDatabase2023}
{AVID}. 2023.
\newblock {{AI Vulnerability Database}}.
\newblock https://avidml.org/.

\bibitem[{Barari, Lucas, and Munger(2021)}]{barariPoliticalDeepfakesAre2021}
Barari, S.; Lucas, C.; and Munger, K. 2021.
\newblock Political {{Deepfakes Are As Credible As Other Fake Media And}} ({{Sometimes}}) {{Real Media}}.
\newblock Preprint, Open Science Framework.

\bibitem[{Bureau(2022)}]{bureau_economic_2022}
Bureau, U.~C. 2022.
\newblock Economic {Census}: {NAICS} {Codes} \& {Understanding} {Industry} {Classification} {Systems}.
\newblock Section: Government.

\bibitem[{Cahlan(2020)}]{cahlan_analysis_2020}
Cahlan, S. 2020.
\newblock Analysis {\textbar} {How} misinformation helped spark an attempted coup in {Gabon}.
\newblock \emph{Washington Post}.

\bibitem[{Christopher(2020)}]{christopherWeVeJust2020}
Christopher, N. 2020.
\newblock We've {{Just Seen}} the {{First Use}} of {{Deepfakes}} in an {{Indian Election Campaign}}.
\newblock \emph{Vice}.

\bibitem[{Cook(2019)}]{cookDeepFakesCould2019}
Cook, E. 2019.
\newblock Deep Fakes Could Have Real Consequences for {{Southeast Asia}}.

\bibitem[{Dao et~al.(2022)Dao, W, Walla, VocalFan, Rubiel, Schreiber, Jaworski, Liu, Williams, Pawlowski, Ammanamanchi, Stadlmann, K{\"u}hne, Taiz, Diethe, {jvmncs}, and {twsl}}]{daoDaviddaoAwfulaiAwful2022}
Dao, D.; W, H.; Walla, A.-A.; VocalFan; Rubiel, E.; Schreiber, F.; Jaworski, J.; Liu, L.; Williams, N.; Pawlowski, N.; Ammanamanchi, P.~S.; Stadlmann, S.; K{\"u}hne, S.; Taiz; Diethe, T.; {jvmncs}; and {twsl}. 2022.
\newblock Daviddao/Awful-Ai: {{Awful AI}} - 2021 {{Edition}}.
\newblock Zenodo.

\bibitem[{De~Bruycker(2017)}]{debruyckerFramingAdvocacyResearch2017}
De~Bruycker, I. 2017.
\newblock Framing and Advocacy: A Research Agenda for Interest Group Studies.
\newblock \emph{Journal of European Public Policy}, 24(5): 775--787.

\bibitem[{{European Commission}(2021)}]{europeancommissionProposalRegulationEuropean2021}
{European Commission}. 2021.
\newblock Proposal for a Regulation on a {{European}} Approach for Artificial Intelligence.
\newblock Technical Report COM(2021) 206 final, European Commission, Brussels, Belgium.

\bibitem[{Goldstein et~al.(2023)Goldstein, Sastry, Musser, DiResta, Gentzel, and Sedova}]{goldsteinForecastingPotentialMisuses2023}
Goldstein, J.~A.; Sastry, G.; Musser, M.; DiResta, R.; Gentzel, M.; and Sedova, K. 2023.
\newblock Forecasting {{Potential Misuses}} of {{Language Models}} for {{Disinformation Campaigns}}---and {{How}} to {{Reduce Risk}}.
\newblock Technical report, Brookings Institution.

\bibitem[{Groh et~al.(2022)Groh, Epstein, Firestone, and Picard}]{grohDeepfakeDetectionHuman2022}
Groh, M.; Epstein, Z.; Firestone, C.; and Picard, R. 2022.
\newblock Deepfake Detection by Human Crowds, Machines, and Machine-Informed Crowds.
\newblock \emph{Proceedings of the National Academy of Sciences}, 119(1): e2110013119.

\bibitem[{Hao(2023)}]{haoChinaPioneerRegulating2023}
Hao, K. 2023.
\newblock China, a {{Pioneer}} in {{Regulating Algorithms}}, {{Turns Its Focus}} to {{Deepfakes}}.
\newblock \emph{Wall Street Journal}.

\bibitem[{Klar(2023)}]{klarDemocratsLaudFederal2023}
Klar, R. 2023.
\newblock Democrats Laud Federal Agency's {{AI}} Move.
\newblock \emph{The Hill}.

\bibitem[{Matthes(2009)}]{matthesWhatFrameContent2009}
Matthes, J. 2009.
\newblock What's in a {{Frame}}? {{A Content Analysis}} of {{Media Framing Studies}} in the {{World}}'s {{Leading Communication Journals}}, 1990-2005.
\newblock \emph{Journalism \& Mass Communication Quarterly}, 86(2): 349--367.

\bibitem[{McGregor(2021)}]{mcgregorPreventingRepeatedReal2021}
McGregor, S. 2021.
\newblock Preventing {{Repeated Real World AI Failures}} by {{Cataloging Incidents}}: {{The AI Incident Database}}.
\newblock \emph{Proceedings of the AAAI Conference on Artificial Intelligence}, 35(17): 15458--15463.

\bibitem[{McGregor, Paeth, and Lam(2022)}]{mcgregorIndexingAIRisks2022}
McGregor, S.; Paeth, K.; and Lam, K. 2022.
\newblock Indexing {{AI Risks}} with {{Incidents}}, {{Issues}}, and {{Variants}}.
\newblock arXiv:2211.10384.

\bibitem[{Montanaro(2022)}]{montanaro_truth_2022}
Montanaro, D. 2022.
\newblock The truth about political ads: {They} can include lies.
\newblock \emph{NPR}.

\bibitem[{Nehamas(2023)}]{nehamas_desantis_2023}
Nehamas, N. 2023.
\newblock {DeSantis} {Campaign} {Uses} {Apparently} {Fake} {Images} to {Attack} {Trump} on {Twitter}.
\newblock \emph{The New York Times}.

\bibitem[{{Partnership on AI}(2023)}]{partnershiponaiPAIResponsiblePractices2023}
{Partnership on AI}. 2023.
\newblock {{PAI}}'s {{Responsible Practices}} for {{Synthetic Media}}.
\newblock Technical report, Partnership on AI.

\bibitem[{Poria et~al.(2017)Poria, Cambria, Hazarika, Majumder, Zadeh, and Morency}]{poriaContextDependentSentimentAnalysis2017}
Poria, S.; Cambria, E.; Hazarika, D.; Majumder, N.; Zadeh, A.; and Morency, L.-P. 2017.
\newblock Context-{{Dependent Sentiment Analysis}} in {{User-Generated Videos}}.
\newblock In \emph{Proceedings of the 55th {{Annual Meeting}} of the {{Association}} for {{Computational Linguistics}} ({{Volume}} 1: {{Long Papers}})}, 873--883. Vancouver, Canada: Association for Computational Linguistics.

\bibitem[{Rana et~al.(2022)Rana, Nobi, Murali, and Sung}]{ranaDeepfakeDetectionSystematic2022}
Rana, M.~S.; Nobi, M.~N.; Murali, B.; and Sung, A.~H. 2022.
\newblock Deepfake {{Detection}}: {{A Systematic Literature Review}}.
\newblock \emph{IEEE Access}, 10: 25494--25513.

\bibitem[{Schegloff(1997)}]{schegloffWhoseTextWhose1997}
Schegloff, E.~A. 1997.
\newblock Whose {{Text}}? {{Whose Context}}?
\newblock \emph{Discourse \& Society}, 8(2): 165--187.

\bibitem[{Schiff, Schiff, and Bueno(2023)}]{schiffLiarDividendImpact2023}
Schiff, K.~J.; Schiff, D.~S.; and Bueno, N. 2023.
\newblock The {{Liar}}'s {{Dividend}}: {{The Impact}} of {{Deepfakes}} and {{Fake News}} on {{Trust}} in {{Political Discourse}}.
\newblock \emph{SocArXiv}.

\bibitem[{Semetko and Valkenburg(2000)}]{semetko2000FramingEuropeanpolitics}
Semetko, H.~A.; and Valkenburg, P. M.~V. 2000.
\newblock Framing {European} politics: {A} {Content} {Analysis} of {Press} and {Television} {News}.
\newblock \emph{Journal of Communication}, 50(2): 93--109.

\bibitem[{Shanahan, McBeth, and Hathaway(2011)}]{shanahanNarrativePolicyFramework2011}
Shanahan, E.~A.; McBeth, M.~K.; and Hathaway, P.~L. 2011.
\newblock Narrative Policy Framework: The Influence of Media Policy Narratives on Public Opinion.
\newblock \emph{Politics \& Policy}, 39(3): 373--400.

\bibitem[{Sharma et~al.(2023)Sharma, Jain, Behl, Baabdullah, Giannakis, and Dwivedi}]{sharmaExaminingMotivationsSharing2023}
Sharma, I.; Jain, K.; Behl, A.; Baabdullah, A.; Giannakis, M.; and Dwivedi, Y. 2023.
\newblock Examining the Motivations of~Sharing Political Deepfake Videos: The Role of Political Brand Hate and Moral Consciousness.
\newblock \emph{Internet Research}, ahead-of-print(ahead-of-print).

\bibitem[{Simonite(2022)}]{simoniteZelenskyDeepfakeWas2022}
Simonite, T. 2022.
\newblock A {{Zelensky Deepfake Was Quickly Defeated}}. {{The Next One Might Not Be}}.
\newblock \emph{Wired}.

\bibitem[{Sorkin et~al.(2023)Sorkin, Warner, Kessler, Merced, Hirsch, and Livni}]{sorkin_i-generated_2023}
Sorkin, A.~R.; Warner, B.; Kessler, S.; Merced, M. J. d.~l.; Hirsch, L.; and Livni, E. 2023.
\newblock An {A}.{I}.-{Generated} {Spoof} {Rattles} the {Markets}.
\newblock \emph{The New York Times}.

\bibitem[{Stokols and Egan(2023)}]{stokolsWhiteHousePress2023}
Stokols, E.; and Egan, L. 2023.
\newblock White {{House}} Press Shop Adjusts to Proliferation of {{AI}} Deep Fakes.
\newblock \emph{POLITICO}.

\bibitem[{Vaccari and Chadwick(2020)}]{vaccariDeepfakesDisinformationExploring2020}
Vaccari, C.; and Chadwick, A. 2020.
\newblock Deepfakes and {{Disinformation}}: {{Exploring}} the {{Impact}} of {{Synthetic Political Video}} on {{Deception}}, {{Uncertainty}}, and {{Trust}} in {{News}}.
\newblock \emph{Social Media + Society}, 6(1): 2056305120903408.

\bibitem[{Williams(2023)}]{williamsExploringLegalApproaches2023}
Williams, K. 2023.
\newblock Exploring {{Legal Approaches}} to {{Regulating Nonconsensual Deepfake Pornography}}.
\newblock \emph{Tech Policy Press}.

\bibitem[{Wittenberg et~al.(2021)Wittenberg, Tappin, Berinsky, and Rand}]{wittenbergMinimalPersuasiveAdvantage2021}
Wittenberg, C.; Tappin, B.~M.; Berinsky, A.~J.; and Rand, D.~G. 2021.
\newblock The (Minimal) Persuasive Advantage of Political Video over Text.
\newblock \emph{Proceedings of the National Academy of Sciences}, 118(47).

\end{thebibliography}

\end{document}